\documentstyle[aps,preprint,eqsecnum]{revtex}
\begin{document}
\draft
\title{
Fractal Spectrum of a Quasi-Periodically Driven Spin System.}
\author{Italo Guarneri\cite{add} and Maria Di Meo }
\address{
Universit\`a di Milano, sede di Como, via Lucini 3,\\
22100 Como, Italy }
\maketitle
\begin{abstract}
We numerically perform a spectral analysis of a quasi-periodically driven
spin 1/2 system, the spectrum of which is Singular Continuous.
We compute fractal dimensions of spectral measures and discuss their
connections with the time
behaviour of various dynamical quantities, such as the moments of the
distribution of the wave packet.  Our data suggest a close similarity
between the information dimension of the spectrum and the exponent ruling the
algebraic growth of the 'entropic width' of wavepackets.
\end{abstract}

\section{Introduction}

The increasingly frequent apparition of Singular Continuous (S.C.) spectra
in various quantum mechanical situations has attracted attention to the
dynamical implications of such spectra. For example, for an electron moving
in an incommensurate or disordered structure, singular continuous spectra
typically result in a sort of pseudo-diffusive dynamics, which has a direct
bearing on transport properties. If the spectrum is a multifractal, some of
these properties depend on the value of certain fractal dimensions,
according to quantitative rules, the precise assessment of which is an
important theoretical task .

SC spectra have also been found in some periodically or quasi-periodically
driven model systems, which were introduced in order to investigate the
possibility of chaotic behaviour in quantum dynamics. \cite{Bel}\cite{KH}.
In particular, quasi-periodically driven spin systems have been studied, in
view of their formal simplicity. Various spectral types have been identified
, on varying levels of evidence, for different systems in this class\cite
{French}. From the mathematical viewpoint, some  quasi-periodically driven
spin systems formally belong to
a class of abstract dynamical systems, which is well-known in ergodic theory,
and for some of which SC spectra have been rigorously
proven to occur\cite{Queff}. In a recent paper \cite{Piko} a renormalization
group analysis has been implemented, strongly supporting SC spectra for a
wider class of spin systems than encompassed by available exact results. The
formal simplicity of this class of systems makes a numerical analysis of their
dynamics particularly convenient, so that they appear very well suited for
the study of the dynamical implications of fractal spectra.

In this paper we describe a numerical spectral analysis for a particular
model in this class, aimed at computing certain fractal dimensions, and at
connecting them to asymptotic aspects of the dynamics. We use a technical
approach, based on a discrete-time variant of Floquet theory, which has
 not been implemented before in this context, and which offers a twofold
advantage. In the first place, if combined with a suitable scheme of
rational approximation for an incommensuration parameter, it allows for a
reliable computation of the spectral measures, at a relatively low
computational cost. Second, the Floquet formulation allows for a meaningful
analysis of the growth in time of the spread of wave-packets, which is a
very efficient empirical marker for continuous spectra, and usually a more
convenient one than the hitherto used decay of correlations.

We then analyze a number of dynamical features, related to the growth of
wave-packets. We find that the exponents of growth of the moments of the
wave packet are not directly related to the Hausdorff dimension of the
spectrum, (which is 1, as shown by a simple rigorous argument). We also find
that the exponent of growth of the ''entropic'' spread of the wavepacket is
close to the numerically computed information dimension of the spectral
measure. Finally, we shortly present a general bound for  the growth of the
information contents
of a
finite string of observations with the length of the
string, in terms of the information dimension of the spectrum.

\section{The Model}

Our model is a periodically kicked spin $\frac 12$ system, with the kicking
strength depending quasi-periodically on time. We consider a function $\hat
S(\varphi )$ from $\left[ 0,2\pi \right] $ into the unitary, unimodular $%
2\times 2$ matrices, given by:

\begin{equation}
\label{S}\hat S(\varphi )=e^{ik\chi (\varphi )\hat \sigma _x}e^{i\hat \sigma
_z}
\end{equation}
where $\chi $ is a periodic real-valued function, to be specified later, $k$
is a parameter, and $\hat \sigma _x,\hat \sigma _z$ are Pauli matrices. In $%
\left[ 0,2\pi \right] $ we consider the shift $\tau $$_\alpha :\varphi
\longmapsto \varphi +2\pi \alpha (mod2\pi ),$ with $\alpha $ a fixed
parameter. We define the dynamics of our model by first arbitrarily fixing a
phase $\varphi _0$, and then prescribing that spinors $\vec \psi (t)=\left|
\begin{array}{c}
\psi _1(t) \\
\psi _2(t)
\end{array}
\right| $ given at (discrete) time $t$ evolve at time $t+1$ into:
\begin{equation}
\label{evol}\vec \psi (t+1)=\hat S(\tau _\alpha ^t\varphi _0)\vec \psi (t)
\end{equation}
At fixed $\varphi _0,\alpha ,k$ the evolution of a given initial spinor is
thus obtained by applying a sequence of unitary matrices, which can be
either a periodic or a quasi-periodic one, depending on the arithmetic
nature of the number $\alpha :$if the latter is a rational, $\alpha =p/q$
with $p,q$ mutually prime integers, the sequence is periodic with period $q,$
otherwise it is quasi-periodic. We have chosen the function $\chi $ as the
periodicized characteristic function of an interval $I$ in $\left[ 0,2\pi
\right] :$ $\chi (\varphi )=1$ if $\varphi $$(mod2\pi )\in I$, $\chi
(\varphi )=0$ otherwise$.$ As we shall explain below, such a choice makes
the numerical analysis very efficient. The sequence of unitary matrices
defining the evolution is now uniquely defined by the symbolic trajectory of
the chosen $\varphi _0$ associated with the shift $\tau _\alpha $ and with
the partition of $\left[ 0,2\pi \right] $ defined by $I$ and its complement.
Since this partition consists of two sets, the symbolic sequence can be
written as a binary sequence. In particular, if the length of the interval $I$
is taken
$2\pi\alpha$ and $\alpha$ , and
$\alpha$ is the (inverse) Golden Ratio $(\sqrt(5)-1)/2$, one obtains the
Fibonacci sequence.

\section{Spectral Measures.}

There are two possible approaches to the spectral analysis of the system (%
\ref{S}). The first consists in considering (\ref{S}) as a classical
dynamical system, the state of which is defined by a couple $\phi ,\vec \psi
$ where $\phi $ is an angle and $\vec \psi $ is a normalized spinor; the
one-step evolution of the system is given by:

\begin{equation}
\label{skew}
(\varphi ,\vec \psi )\longmapsto (\tau _\alpha \varphi ,\hat S(\varphi )\vec
\psi )
\end{equation}

The system thus defined belongs to a class of dynamical systems known as
"skew-products"\cite{Queff}. Spectral measures are defined from the Fourier
transform of correlation functions; in particular, given an initial spinor $%
\vec\psi(0)$, the corresponding spectral measure $d\mu$ is defined by:
\begin{equation}
\label{ftransf}R(t)=\int\limits_0^{2\pi }e^{it\lambda }d\mu (\lambda )
\end{equation}
where $R(t)$ is the correlation function:
\begin{equation}
\label{corr}R(t)=\lim _{T\rightarrow \infty }\frac
1T\sum\limits_{s=0}^T\left\langle \vec \psi (s),\vec \psi (s+t)\right\rangle
\end{equation}
where $\left\langle ,\right\rangle $ is the scalar product in $C^2.$
Spectral measures defined in this way will be termed "dynamical" measures in
the following.

Another definition of spectral measure rests on a generalization of Floquet
theory, which is obtained on imbedding the nonautonomous quantum dynamics
\ref{evol} in an autonomous dynamics, defined in a larger Hilbert space.
This is done as follows. We consider the phase $\varphi $ as a new dynamical
variable, and thereby consider state vectors $\Psi =$$\vec \psi (\varphi )$
in the Hilbert space ${\cal H}=L^2(\left[ 0,2\pi \right] )\otimes C^2$. In
this space we consider the discrete unitary group generated by the unitary
Floquet operator:
\begin{equation}
\label{floq}({\cal S}\vec \psi )(\varphi )=\hat S(\varphi )\vec \psi (\tau
_\alpha ^{-1}\varphi )
\end{equation}
If $\vec\psi(\varphi)$ is regarded as a curve in the phase space of the
dynamical system (\ref{skew}), then (\ref{floq}) specifies the evolution of
this curve under the dynamics (\ref{skew}).
Our second definition of a spectral measure is just the usual one for unitary
operators, applied to the operator (\ref{floq}).
Such spectral measures will be termed "Floquet measures".
The connection between dynamical and Floquet measures will be discussed
separately for the incommensurate and the commensurate cases. For the time
being, we shall point out a simple property of the spectrum of ${\cal S}$,
which is valid in both cases. On defining in ${\cal H}$ the unitary operator
${\cal U}\vec \psi (\varphi )=e^{i\varphi }\vec \psi (\varphi )$ we find:%
$$
{\cal U}^{\dagger }{\cal SU}=e^{-2\pi i\alpha }{\cal  S}
$$
which implies that the spectrum of ${\cal S}$ is invariant under the shift $%
\tau _\alpha $ : that is, if a point of the unit
circle belongs in the spectrum of ${\cal S}$, so does the whole orbit of that
point under $\tau _\alpha $ .

It is worth remarking that the operator ${\cal S}$ can be interpreted as the
Floquet operator for a linear kicked rotator endowed with spin; the
occurrence of SC spectra for a spinless linear kicked rotator has been
discussed in \cite{Bel}.

\section{The commensurate case.}

In the commensurate case $\alpha =p/q,$ the model is a periodically driven
one, and its dynamics can be understood, as usual, from a spectral analysis
of the one-period evolution operator. The state vector at times multiple of
the period $q$ is found from:%
$$
\vec \psi (nq)=\hat T^n(\varphi _0)\vec \psi (0)
$$
where
\begin{equation}
\label{T}\hat T(\varphi _0)=\prod\limits_{t=0}^{q-1}\hat S(\tau _\alpha
^t\varphi _0)
\end{equation}
(the product being ordered from right to left) yields the evolution over one
period. This operator does not depend on time; thus the dynamics, read at
integer multiples of the period, is given by a discrete unitary group,
generated by a fixed unitary (\ref{T}). Being a finite matrix, the latter
has obviously a pure discrete spectrum, consisting of two conjugate
eigenvalues $z_1=e^{i\lambda },$ $z_2=z_1^{*}=e^{-i\lambda }.$ The dynamics
is therefore recurrent.

{}From (\ref{floq})(\ref{T}) we find:
\begin{equation}
\label{fiber}({\cal S}^q\vec \psi )(\varphi )=\hat T(\varphi )\vec \psi
(\varphi )
\end{equation}
that is, ${\cal S}^q$ is a fibered operator. Its fibers are precisely the $%
2\times 2$ matrices$\hat T(\varphi ),$ and its spectrum is the range of the
functions $z_1(\varphi ),z_2(\varphi )$ giving the eigenvalues of $\hat
T(\varphi ).$ It turns out that, when the size of the interval $I$ is not a
multiple of $2\pi /q,$ both functions have exactly two values in their
range: otherwise, they have just one. This fact allows for a decisive
simplification of the numerical analysis, and is proved in Appendix A; it
stems from the finiteness of the range of the function $\chi .$

Therefore the spectrum of ${\cal S}^q$ consists of a finite number $2M$ of
eigenvalues $z_N=e^{i\Lambda _N},(N=1,...2M)$, in complex conjugate pairs,
with $M=2$ or $M=1.$ The spectrum of ${\cal S}$ is then a subset of the set
of the $2Mq$ complex $q$-th roots of the $z_N$. The corresponding
eigenphases are:
\begin{equation}
\label{eigen}\lambda _{j,N}=\frac{\Lambda _N}q+\frac{2\pi j}q
\end{equation}
The spectrum of ${\cal S}$ is discrete, i.e., the Floquet dynamics in the
extended Hilbert space has the same spectral character as the proper
dynamics. It should be noted that
 the fact that the dynamical measure and the Floquet one are of
the same type is, in the commensurate case,  a nongeneric feature , connected
with the choice of a
finitely valued function $\chi $ in (\ref{evol}). On choosing a nonconstant
analytic function $\chi $, an absolutely continuous spectrum of ${\cal S}$
would be found instead, in contrast to the character of the dynamics, that
would be still recurrent.

\section{The incommensurate case.}

The spectral measure of a vector $\Psi =$$\vec \psi (\varphi )$ with respect
to the operator (\ref{floq}) is defined as the Fourier transform of the
correlation:
\begin{equation}
\label{corr2}C(t)=\left\langle \Psi ,{\cal S}^t\Psi \right\rangle _H=\int
d\varphi \left\langle \vec \psi (\varphi ),S(\varphi )S(\tau _\alpha
^{-1}\varphi )...S(\tau _\alpha ^{-t}\varphi )\vec \psi (\tau _\alpha
^{-1-t}\varphi )\right\rangle
\end{equation}
In the incommensurate case, the well-known ergodic property holds:
\begin{equation}
\label{ergo}\frac 1{2\pi }\int d\varphi g(\varphi )=\lim _{T\rightarrow
\infty }\frac 1T\sum\limits_{s=0}^Tg(\tau _\alpha ^{-s}\varphi _0)
\end{equation}
for any summable function $g,$and for almost every $\varphi _0.$ Then the
integral over $\varphi $ in (\ref{corr2}) can be computed as follows:
\begin{equation}
\label{corr3}C(s)=\lim _{T\rightarrow \infty }\frac
1T\sum\limits_{t=0}^T\left\langle \vec \psi (\tau _\alpha ^{-t}\varphi
_0),\hat S(\tau _\alpha ^{-t}\varphi _0)\hat S(\tau _\alpha ^{-t-1}\varphi
_0)...\hat S(\tau _\alpha ^{-t-s}\varphi _0)\vec \psi (\tau _\alpha
^{-1-t-s}\varphi _0)\right\rangle
\end{equation}
Now we apply to both factors in the scalar product the unitary operator $\hat
S(\varphi _0)\hat S(\tau _\alpha ^{-1}\varphi _0)...\hat S(\tau _\alpha
^{-t+1}\varphi _0)$ and take $\vec\psi=\vec\psi(0)$. Then, on
comparing the result with (\ref{corr}) we find that

$%
C(t)=R(t)$ except possibly for a set of zero measure of values of $\varphi
_0.$ Therefore, in the incommensurate case the dynamical spectral measure (%
\ref{corr}) coincides with the Floquet spectral measure.

In \cite{Piko} a quite similar quasi-periodically kicked spin system was
studied, which was reduced to a classical dynamical system for which
singular continuity of the spectrum is a known mathematical result. It looks
likely, on account of the closeness of our model to that one, that this result
holds in our case, too; our numerical results, presented below, yield strong
evidence in this sense.

Since in the incommensurate case the orbit of any point of the unit circle
under the shift is dense in the circle, from the invariance of the spectrum
under $\tau _\alpha $ it follows that the spectrum (which is by definition a
closed set) must coincide with the whole unit circle.

\section{Numerical Results}

Our main focus is on the incommensurate case, with $\alpha $ given by the
Golden
Ratio. In most of our computations we have taken $I$ an interval of length
$2\pi /3$; in a few cases we have taken $2\pi\alpha$ instead. Unless
explicitly stated, we will make reference to the first choice.
We have
obtained two types of numerical data: dynamical ones, obtained from
directly simulating the Floquet dynamics, and spectral ones. We discuss the
latter first.

A direct computation of the spectral measure from the Fourier transform of
the numerically computed correlation function (\ref{corr}) involves certain
subtleties, well-known in the spectral analysis of time series, connected
with the necessity of appropriately weighting  the tails of
correlation functions\cite{ABC}. A more convenient strategy for the present
model is based on approximations of the Golden Ratio constructed via
continued fraction expansion, which produces a well-known sequence of
rational approximants. For such approximants we carefully compute the
spectral measures, as we explain in Appendix B, and thus we obtain
approximations of the true spectral measure, by means of the pure point
measure (\ref{mis}).

A similar strategy has been widely used in the numerical investigation of
other incommensurate spectral problems, such as e.g. the Harper and the
Kicked-Harper model\cite{KH}. It is worth remarking, however, that in those
cases, unlike the present one, the commensurate approximations have an {\it %
absolutely continuous} band spectrum, that is usually taken as a covering of
the limit spectrum.

For the obtained measures we have performed a fractal analysis, aimed at the
determination of the fractal dimensions $D_{q}$; the Hausdorff dimension $D_H
$ is $1$ in the incommensurate case, because the spectrum in that case is
the whole unit circle. In our computations we have used the definitions,

\begin{equation}
\label{dim}D_1=\lim _{\delta \rightarrow 0}\frac{\sum_i\mu _i\ln \mu _i}{\ln
\delta };\qquad D_{q}=\lim _{\delta \rightarrow 0}\frac{\ln \sum_i\mu _i^q}{
(q-1) \ln \delta }
\end{equation}
where the interval $\left[ 0,2\pi \right] $ was partitioned into small
intervals of size $\delta $, the $i-$th of which receives a weight $\mu _i$
from the spectral measure.

$D_q$ were obtained by a linear fit of the numerators in (\ref{dim}) vs the
denominators in a suitable range of $\delta $. A typical result is shown in
fig.1. In no case could the value  $D_{0\text{ }}$obtained from the spectrum
of generalized dimensions  be distinguished from $D_H=1.$ The correlation
dimension  $D_2$ is known to rule the decay of integrated correlations
\cite{Geis2}:

\begin{equation}
\label{d2}\frac 1T\sum\limits_{t=0}^{T-1}\left| C(t)\right| ^2\sim
const.\cdot T^{-D_2}
\end{equation}
asymptotically for $T\rightarrow \infty .$ We have used this exact result to
adjust our numerical computation of the other dimensions; in fact, since the
approximating measure is a pure-point one, the dimensions of the limit measure
must be read off a
suitable range of not too small $\delta $. Our appreciation of a reliable
range for $\delta $ was based on the comparison of the rhs of the 2nd eq.(%
\ref{dim}) with the exponent $\gamma $ of algebraic decay of correlation
obtained from a direct computation of the latter (Fig. 2).

We have also computed a number of dynamical data, by a direct numerical
simulation of the dynamics. In particular, we have analyzed the
growth in time of the momenta $I_{\alpha}$ of the distribution of the
Floquet wave-function on the Fourier basis. $I_{\alpha}$ are defined by:
\begin{equation}
\label{mom}
\begin{array}{c}
I_{\alpha}(t)=\sum_{n=-\infty }^{+\infty }\left| n\right| ^{\alpha}p_n(t) \\
p_n(t)=\frac 1{2\pi t}\sum\limits_{s=0}^{t-1}\sum\limits_{j=1}^2\left|
\int\limits_0^{2\pi }d\varphi e^{in\varphi }\psi _j(\varphi ,s)\right| ^2
\end{array}
\end{equation}
Notice that moments of order $\alpha \geq 1$ diverge for $t>0$, because the
Fourier coefficients of the wave function behave as $n^{-1}$ at large $n$,
due to the discontinuity of $\chi$ .

Finally we have studied the growth in time of the average entropy $S(t):$

$$
S(t)=-\sum_{n=-\infty }^{+\infty }p_n(t)\ln p_n(t)
$$

Examples of the dependence on time of the moments and of the entropy are
given in figs.3,4. The moments increase according to a power law: $I_\alpha
(t)\sim const.t^{\alpha \beta (\alpha )}$, and so does the ''entropic number
of states'': $N(t)=\exp (S(t)\sim const.t^\sigma $. In Table 1 we summarize
the values of the exponents $\gamma ,$$\beta ,\sigma $ and the values of the
$D_1,D_2$ for several choices of the kicking strength $k$.

The specified errors are those involved in linear fits of bilogarithmic
plots used to compute fractal dimensions or growth exponents; they do not
include numerical errors in computing the data themselves, therefore
 they to some extent
underestimate the real errors . Generally speaking, it is usually difficult to
get precise estimates for the dynamical growth exponents associated with
fractal spectra,  partly because of finite-basis
effects coming into play at large times, and much more because the curves of
growth display a characteristic pattern of kinks . In cases in which such
seemingly log-periodic structures were particularly evident, the growth
exponent was obtained directly by drawing a straight line through a sequence
of maxima. In such cases the fitting error, not specified in Table 1, is on
the order of the first missing digit.

\section{Discussion.}

Some general exact results are known about the long-time properties of the
quantum dynamics in the presence of a SC spectrum. First of all, integrated
correlations like (\ref{d2}) must tend to zero, and momenta $I_\alpha $ must
diverge in the limit $t\to \infty $ in all cases when the spectrum is purely
continuous; furthermore, if the spectral measure is a fractal one, with a
correlation dimension $D_2$, and information dimension $D_1$, then (\ref{d2}%
) holds for the decay of correlations\cite{Geis2}, and $\beta (\alpha )\geq
D_1$\cite{IG}. Data in Table 1 are fully consistent with these exact
estimates (as mentioned above, the estimate (\ref{d2}) was actually used to
adjust our numerical method). In addition,  more or less heuristic arguments
relating more precisely the exponents $\beta (\alpha )$ to multifractality
have been attempted. One such argument \cite{Geis1} has led to hypothesize that
$\beta
(\alpha )=D_H$, the Hausdorff dimension of the spectrum. Although consistent
with numerical results from a number of models, this hypothesis has been
called in question \cite{Wil}, and also recent numerical investigations have
provided evidence that $\beta (\alpha )$ is not in general a constant but
covers a continuous range of scaling exponents (multi-scaling). Our present
data are fully consistent with the latter picture, because the observed
values of $\beta (\alpha )$ differ from $D_H=1$ (which was theoretically
established and numerically confirmed) significantly more than the estimated
numerical error. Somewhat smaller, but still significant, are the differences
observed in
the values of $\beta (\alpha )$ obtained for different values of $\alpha $.

Two interesting facts emerging from table 1 are (i) the closeness of the
values obtained for different values of $k$, which seems to indicate that the
fractal structure of the spectrum is essentially determined by the
quasi-periodic structure
of the symbolic strings alone, and (ii) the closeness of the
value of  the "entropic" exponent $\sigma$ to that of $D_1$.
Although the difference
of the two values was comparable to the relatively large numerical error,
the actual agreement may be much better, because many of the data used in
estimating
$D_1$ appear not to have fully converged to the proper asymptotic regime.
These data are in fact compared to the dynamical entropic exponent in Fig.5 ;
the
agreement "by eye" is there better than implied by Table 1.
 Whether this fact
reflects a real connection between
the two quantities is an interesting theoretical question, for which we have
no answer for the time being.

Finally we shall mention a dynamical property, that, although not directly
related to the above numerical results, has been recently
introduced in essentially the same context as discussed here, with the aim
of analyzing the possibly chaotic property of the evolution of the spin
system. For a classical dynamical system in discrete time, with a compact
phase space $\Omega $, the ''informational complexity'' is introduced by
considering a finite partition $\Pi $ consisting of $m$ subsets,
and the associated symbolic dynamics.
For any given integer time $t$, all the possible symbolic strings of length $%
t$ define a partition $\Pi ^t$ of $\Omega ^t$. If a specific orbit of the
system is chosen, then a probability can be attached to every class of this
partition, defined as the frequency with which the given finite string
appears in the infinite symbolic string of the given trajectory. The Shannon
entropy $H(t)$ can then be defined, as well as the corresponding number of
histories $N(t)=\exp (H(t))$. A chaotic behaviour is associated with an
exponential growth of $N(t)$ in time.

In a recent paper \cite{Rom} such an  analysis has been numerically
implemented for  the case of a quasi-
periodically driven spin system. In essence,  $\Omega$ was taken as the
compact variety of normalized spinors;  a certain partition
of $\Omega$ was constructed, and  the growth of $H(t)$ with time was
analyzed. A seemingly sub-exponential growth of  $%
N(t)$ was observed, of the type $\exp(ct^{\gamma})$ with $\gamma< 1$.

We shall here sketch a general argument, which
establishes an upper bound for $N(t)$ in the presence of a fractal spectrum.
 At time $t$, $\Pi ^t$ is a partition of a $t.\nu $%
-dimensional space, where $\nu $ is the dimension of $\Omega $; therefore $%
H(t)\leq t\nu \log m$. Now, all the strings $\psi
(s+1),...\psi (s+t)$ of length $t$  which are observed  in the evolution do
indeed span a
vector space of dimension $t$, because of the pure continuity of the
spectrum; however, their ''effective dimension'' can be much less. It can in
fact be proven \cite{IG1} that the minimum dimension $d_\epsilon (t)$
of a subspace which contains all the strings of length $t$ within a maximum
error $\epsilon $ asymptotically grows like $t^{D_1}$. This means that,
apart from a small error of order $\epsilon $, the entropy $H(t)$ is the
same as the one computed over $\sim m^{c\nu t^{D_1}}$ classes. This leads
to the upper bound $H(t)\leq const.t^{D_1}$.

Generally speaking, a direct numerical analysis of $N(t)$
 seems difficult, because computing the frequency  of strings
long enough to
reproduce the correct asymptotic regime requires an enormous computation time.
In any case, for the special case of quasi-periodically driven spin systems,
the
above upper bound is but a very crude one: in fact a
simple argument \cite{Ark} indicates  that the asymptotic
growth of $N(t)$ cannot be faster than algebraic.

\section{Appendix A}

The matrix $\hat T(\varphi )$ is uniquely associated with  a binary string
 of $q$ digits, which specifies the symbolic periodic trajectory
of $\varphi .$ Suppose that the strings corresponding to two matrices $\hat
T(\varphi ),\hat T(\varphi ^{\prime })$ differ from each other merely by a
cyclic permutation of digits. Then the two matrices differ from each other
by a cyclic permutation of the $\hat S$ operators which enter as factors in
the definition of the matrices themselves. The two
matrices are then unitarily equivalent, and have therefore the same
eigenvalues.
We have thus reached the conclusion that the number of distinct values in
the range of $z_1(\varphi )$ (and of $z_2(\varphi )$, as well) is equal to
the number of nonequivalent symbolic strings of $q$ digits, two strings
being equivalent if they can be obtained from each other by a cyclic
permutation. Thus in order to find this number we have to find the total
number of symbolic strings of length $q,$ and to divide it by $q$, which is
the number of strings in an equivalence class. Let us denote by $C_0$ and $%
C_1$ (with $C_1=I)$ the two classes of the partition. Then the points $%
\varphi $ which produce a given symbolic string $i_1,i_2,...i_q$ are those
belonging to the set $A_{i_1i_2...i_q}=\tau _\alpha ^{-1}(C_{i_1})\cap
...\cap \tau _\alpha ^{-q}(C_{i_q}).$ The sets $A_{i_1i_2...i_q}$ define a
partition of $\left[ 0,2\pi \right] ,$ and there are as many distinct
symbolic strings of length $q$ as are nonempty classes in this partition. It
is easily seen that there are either $2q$ or $q$ nonempty such classes, the
latter case occurring if, and only if, $I$ is a multiple of $2\pi /q;$ in
fact, on repeatedly applying the shift $\tau _\alpha $ to the interval $I,$
we get $q$ distinct intervals, whose endpoints make a set of $2q$ or $q$
distinct points, depending on whether the length of $I$ is a multiple of $%
2\pi /q,$or not. The sets $A_{i_1i_2...i_q}$ are precisely the disjoint
intervals in which $\left[ 0,2\pi \right] $ is divided by these $2q$ (resp.,
$q$) points, and their number is therefore $2q$ (resp., $q$). The number of
nonequivalent strings is thus exactly $2$ (resp., $1$).

\section{Appendix B}

In a commensurate case with $\alpha =p/q$ we get, from the definition (\ref
{corr2}) ::
\begin{equation}
\label{corr4}C(t,\Psi )=\sum\limits_{N=1}^{2M}e^{it\Lambda
_Nq^{-1}}\sum\limits_{j=0}^{q-1}\left\| \hat P_{j,N}\Psi \right\|
_H^2e^{2\pi itjq^{-1}}
\end{equation}
where $\hat P_{j,N\text{ }}$ is projection over the eigenspace of $\hat c$
corresponding to the eigenphase $\lambda _{j,N}.$ The spectral measure in $%
\left[ 0,2\pi \right] $ is pure point:
\begin{equation}
\label{mis}\sum\limits_{N=1}^{2M}\sum\limits_{j=0}^{q-1}\left\| \hat
P_{j,N}\vec \psi \right\| _H^2\delta (\lambda -\lambda _{j,N})
\end{equation}
In order to compute this measure we have to find the eigenphases $\lambda
_{j,N\text{ }}$ and the associated weights $p_{j,N}=\left\| \hat P_{j,N}\Psi
\right\| _H^2.$ The eigenphases are most easily computed, by diagonalizing
the $2\times 2$ matrix (\ref{T}), for two different values of $\varphi _{0
\text{ }}$(e$.$g., the two extremes of the interval $I$), and then using (%
\ref{eigen})(we assume that $q$ is not a multiple of $3$, for in that case
just one value of $\varphi _0$ would be sufficient.). Concerning the
weights, denoting $\hat P_N=\sum_j\hat P_{j,N},$ we have from (\ref{corr4}):
\begin{equation}
\label{corr5}e^{-it\Lambda _Nq^{-1}}C(t,\hat P_N\Psi
)=\sum\limits_{j=0}^{q-1}p_{j,N}e^{2\pi itjq^{-1}}
\end{equation}
which is an exactly periodic function of the discrete time $t$ with period $%
q $, so that the weights are quite easily and reliably computable via finite
Fourier transform as soon as the lhs of (\ref{corr5}) is known. Since $%
\Lambda _{N\text{ }}$ is computed as described above, we are left with the
computation of the correlation of $\hat P_N\Psi $ at times $t=0,...q-1.$ To
this end we first find $\hat P_N\Psi $ as follows: having diagonalized the
matrix (\ref{T}) for all values of $\varphi _{0\text{ }}$in a suitably thick
grid, we let%
$$
\hat P_N\vec \psi (\varphi _0)=\left\langle \vec u_N(\varphi _0),\vec \psi
(\varphi _0)\right\rangle \vec u_N(\varphi _0)
$$
where $\vec u_N(\varphi _0)$ is defined as follows: it is the eigenvector of
(\ref{T}) with eigenvalue $\Lambda _N,$ if the latter is an eigenvalue of (%
\ref{T}) at the given $\varphi _0,$ and it is the zero vector otherwise. The
correlation of $\hat P_N\Psi $ is then found directly from the definitions (%
\ref{floq}),(\ref{corr2}). The only approximation involved in this
computation is the discretization of the scalar product in ${\cal H}$, which
involves an integral over $\varphi $ and is instead computed as a finite sum
over the chosen finite grid used to discretize $\left[ 0,2\pi \right] $ .

\smallskip\

We thank R.Artuso for valuable advise in computing fractal dimensions,
and A.Vulpiani for an useful discussion.

\section{Figure Captions}

Fig.1 - Spectrum of generalized fractal dimensions $D_q$, for $k=10.5$. The
rational approximant of the Golden Ratio used in this computation was $%
1597/2584$.

Fig.2 - A bilogarithmic plot of $\sum \mu_i^2$ versus $\delta^{-1} $
(eqn.(\ref{dim}%) used for estimating $D_2$.
The slope of the straight line is the exponent
of the correlation decay, eqn.(\ref{d2}). Here $k=10.5$.

Fig.3 - Illustrating the growth of the moment $I_{1/2}$, for $k=4$.

Fig.4 - Illustrating the growth in time of the entropy for $k=10.5$. The
slope of the straight line is the computed information dimension $D_1$.

Fig.5 - A bilogarithmic plot of $-\sum \mu_i\log \mu_i$ versus $\delta^{-1}$
(eqn.(\ref{dim}), for the case $k=1$ of Table I. The slope of the dashed line
is the entropic exponent
$\sigma$, but the position of the line has been chosen arbitrarily.

\begin{table}
 \begin{tabular}{|l|l|l|l|l|l|l|} \hline
{$k$}  & {$\gamma$}  &  {$D_{2}$}  &  {$\sigma$}  & {$D_{1}$}  & {$\beta(1/2)$}
& {$\beta(3/4)$} \\ \hline
 1 &  0.47  & 0.49$\pm$0.01 &0.736$\pm$0.001 & 0.7$\pm$0.11  &0.94 &0.91 \\
 9 &  0.53  & 0.544$\pm$0.005 &0.755$\pm$0.001  &0.7$\pm$0.05   &0.96  &0.9 \\
 16 &  0.54  & 0.549$\pm$0.004 &0.76$\pm$0.002   &0.75$\pm$0.05  &0.94
&0.89\\
110.2 &  0.564 & 0.565$\pm$0.004 &0.76$\pm$0.002  & 0.76$\pm$0.03  &0.94  &0.9
\\ \hline
  \end{tabular}
  \caption{Some Fractal Dimensions and Dynamical exponents . The ratio of the
size of the interval $I$ to $2\pi$ was the Golden Ratio for the 1st row,
$2/3$ for the others.
\label{I}}
\end{table}

\end{document}